\documentclass{INTERSPEECH2023}

\interspeechcameraready

\usepackage{color, colortbl}
\usepackage{marvosym}
\title{A Novel Interpretable and Generalizable Re-synchronization Model for Cued Speech based on a Multi-Cuer Corpus}
\name{Lufei Gao$^1\dagger$, Shan Huang$^1\dagger$\thanks{$\dagger$ Equal contribution. This work was supported by the National Natural Science Foundation of
China (No. 62101351)}, Li Liu$^2$$^{(\textrm{\Letter})}$
}
\address{
  $^1$Shenzhen Research Institute of Big Data, China\\
  $^2$The Hong Kong University of Science and Technology (Guangzhou), China}
\email{avrillliu@hkust-gz.edu.cn}

\begin{document}

\maketitle
 
\begin{abstract}
Cued Speech (CS) is a multi-modal visual coding system combining lip reading with several hand cues at the phonetic level to make the spoken language visible to the hearing impaired. Previous studies solved asynchronous problems between lip and hand movements by a cuer\footnote{The people who perform Cued Speech are called the cuer.}-dependent piecewise linear model for English and French CS. In this work, we innovatively propose three statistical measure on the lip stream to build an interpretable and generalizable model for predicting hand preceding time (HPT), which achieves cuer-independent by a proper normalization. Particularly, we build the first Mandarin CS corpus comprising annotated videos from five speakers including three normal and two hearing impaired individuals. Consequently, we show that the hand preceding phenomenon exists in Mandarin CS production with significant differences between normal and hearing impaired people. Extensive experiments demonstrate that our model outperforms the baseline and the previous state-of-the-art methods.
\end{abstract}

\section{Introduction}

Hearing-impairment is one of the social-concerned issues worldwide. According to the World Health Organization (WHO), more than 1.5 billion people worldwide are affected by some form of hearing impairment \cite{whoint}, presenting significant challenges for both society as a whole and the communities affected by this issue.
\textbf{Cued Speech (CS)} is a visual coding system for the spoken language invented by Cornett in 1967 \cite{cornett1967cued}.
CS differs from sign language (SL) \cite{liddell1989american,valli2000linguistics,stokoe2005sign} in that it follows the rules of a normal language system, using hand gestures to code phonemes and supplementing the limitations of lip reading.

Currently, CS has been adapted to about 65 languages and dialects all over the world \cite{csorg}. 
Liu et al. proposed a Chinese cued speech system, specifically for Mandarin cuers \cite{liu2019pilot,liu2022objective}.
In practice, we have found that for adults either normal hearing or hearing-impaired, who master Chinese \textit{Pinyin}, it only takes 24 hours for them to master the rules of Mandarin CS and use it for simple expressions.
Similar to other languages, hand cues in Mandarin CS also tend to reach the target state before the lip cues during expression.
In the literature, it is called hand preceding phenomenon or lip-hand asynchronous problem \cite{attina2004pilot, attina2006temporal, liu2020re}.
Comprehending and dissecting this phenomenon holds great significance in understanding the cognitive principles of CS expression, improving automatic CS recognition \cite{liu2017automatic,liu2018automatic,liu2018visual}, and developing realistic CS synthesis systems.

There are three main contributions in this work:
\begin{itemize}
    \item A new multi-cuer mandarin CS corpus is developed with accurate manual annotation for the lip and hand movements. It contains CS videos produced by \textcolor{black}{three normal cuers and two hearing loss cuers}. As far as we know, this is the first Mandarin CS corpus that contains data from hearing-loss people which is incredibly valuable for the academic research of CS. 
    \item A novel method is proposed to address the lip-hand multi-modal asynchronous problem in the CS system and to deal with cuer adaptation. The method utilizes three lip stream measures to achieve lip-hand alignment for temporal segmentation and selects proper strategies for cuer normalization. 
    \item  \textcolor{black}{Extensive experimental results show that our proposed re-synchronization method archives better performance compared with the previous methods concerning the re-synchronization effect, interpretability, and generalization ability for multiple cuers. Besides, we discuss the hearing-impaired performance of their mandarin CS expression with some noteworthy observations.}
\end{itemize}


\section{Related Work}

\subsection{Hand preceding phenomenon}
\label{section:preprints}
The temporal organization of hand and speech coordination during French CS production was studied in \cite{attina2004pilot, attina2006temporal}. 
It was found that hand movement reaches the target position 239 ms before the onset of the acoustic vowel and hand shape formation accounts for a large fraction of hand transitions \cite{attina2004pilot}. 
This relationship is not occasional but was also found on other proficient CS cuers \cite{attina2006temporal}. 
The study \cite{aboutabit2006hand} measured a 144.5 ms long HPT for syllables extracted from French continuous sentences and affirmed the importance of the instant at which the hand reaches the target position.

\subsection{Lip-hand re-synchronizations in CS recognition}
To resolve the issue of lip-hand asynchrony, Liu et al. \cite{liu2020re,liu2018automatic,liu2019novel} analyzed the correlation between HPTs and their lip target temporal positions, where a piece-wise linear function is used for HPM.
It is presumed that the HPT remains constant until the turning point is achieved, after which it quickly declines. 
As the model is based on limited English/French CS videos, it is difficult to generalize to new cuers producing Mandarin CS, thus less generalizable.

Asynchronous multi-modal fusion is a fast-growing research area \cite{tang2016integration,sun2019active}.
Prior researches have explored the use of neural networks to tackle this issue in CS.
In \cite{wang2021cross,sankar2022multistream}, RNNs trained with a CTC loss were adopted to address the asynchronous problem. 
In \cite{liu2022cross}, a transformer based on cross-modal mutual learning was proposed by achieving feature fusion based on the aligned modalities with the long-time dependencies.
However, the black-box methods lack interpretability, which limits our ability to identify the specific factors related to the issue.

In the following sections, we will describe the corpus construction and define three measurements on the acoustic stream and employ statistical-based display methods to address the asynchronous issue, thereby enhancing both the interpretability and generalization ability of the multi-modal fusion process.

\section{Methodology}

\subsection{Mandarin CS Corpus}
\subsubsection{Data collection}
To collect the Mandarin CS videos, five volunteers, including three normal hearing (2 female and 1 male, named NF1, NF2, and NM1) and two hearing-impaired (1 female and 1 male, named DF1 and DM1) learned to use Mandarin CS in advance within 24 hours of training. A 1000-sentence Chinese corpus consisting of 5 themes was constructed for normal-hearing cuers, while a smaller corpus consisting of simpler sentences was provided for the hearing impaired cuers.
Each sentence contains 4 to 27 words, 10.6 on average. 
The volunteers recorded 1000 videos at home, using the front-facing cameras on their personal mobile phones with the following standards: 
1) white background with no decoration; 2) 720 HD at 30 fps; 3) turning off the camera mirror effect. 
Finally, we collected 3000 CS videos by the normal and 979 by the hearing-impaired\footnote{All cuers have consented to the public release of the data. The corpus can be accessed from: https://github.com/lufei321/ReSync-CS.}.

\subsubsection{Data annotation}
Each video contains two annotation documents of temporal segments corresponding to phonemes, one based on hand gestures and the other based on acoustics. Hand annotations were completed by ELAN \cite{elan}, while acoustic annotations by Praat \cite{boersma2011praat}. 
The annotations were all manually done and double-checked.

\subsection{Hand preceding analysis}
\begin{figure}[ht]
    \centering 
    \includegraphics[width=\linewidth]{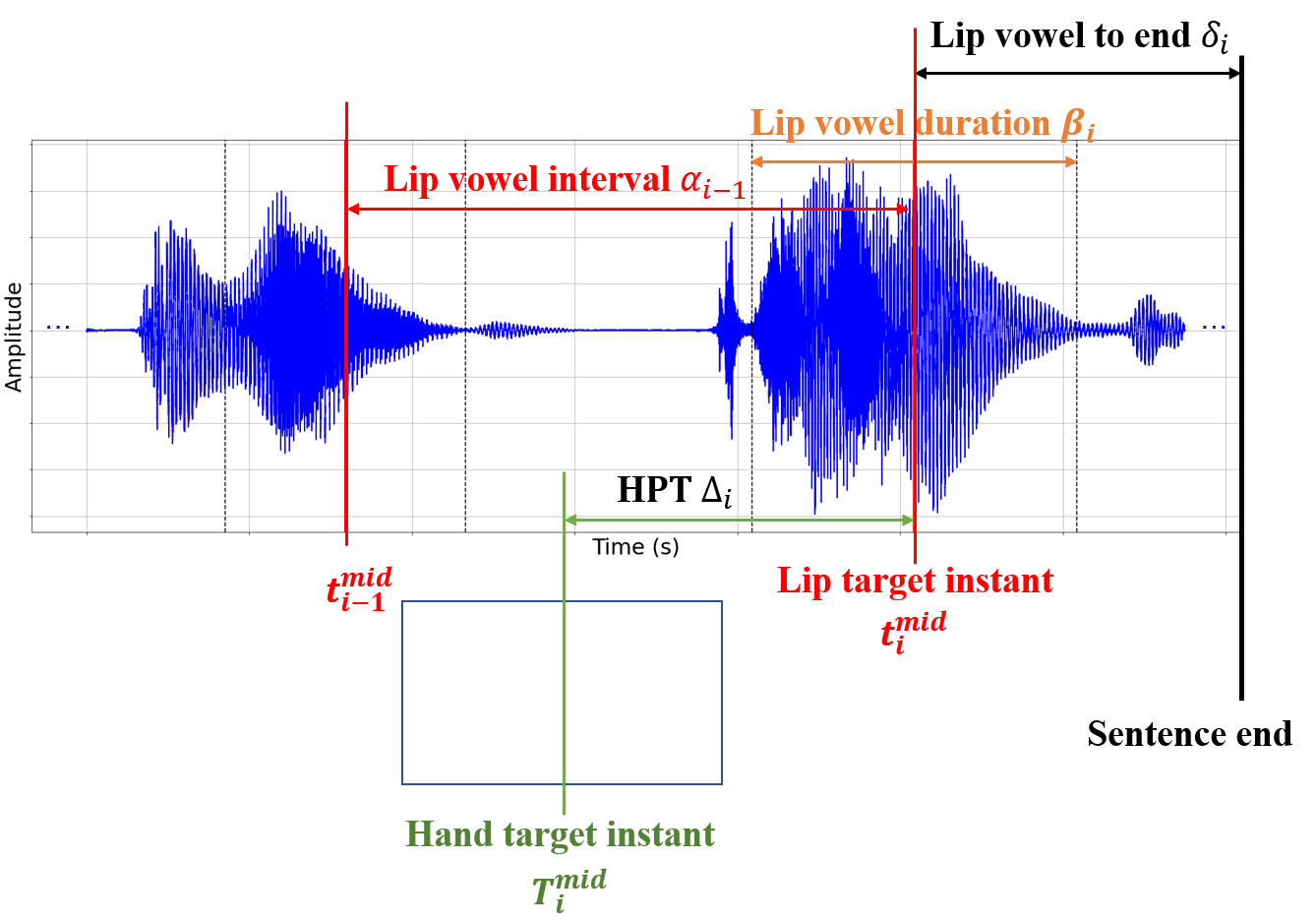} 
    \caption{Hand preceding analysis illustration. \textbf{Green}: the box refers to the hand temporal segment corresponding to the vowel at $t_i^{mid}$; the arrow refers to HPT.  \textbf{Black}: the vertical line is the sentence end; the arrow refers to LVE. \textbf{Red}: two vertical lines are the lip target time instants of two adjacent vowels; the arrow refers to LVI.  \textbf{Orange}: the arrow refers to LVD. } 
    \label{fig:hpm} 
\vspace{-0.5cm}
\end{figure}

For an input video, the audio signal and visual signal are regarded as the lip stream and hand stream, respectively. 
Due to the fact that one vowel corresponds to one syllable in Mandarin, only vowel segments are taken into our consideration.
The required measurements related to the analysis are drawn in Fig. \ref{fig:hpm}.
Specifically, HPT is defined as the time difference between two instants, i.e., the hand target instant $T_i^{mid}$ and lip target instant $t_i^{mid}$ of the $i^{th}$ vowel in a sentence. 
The target instants are the mid-time points between the start and 
the end time of the vowel segments extracted from temporal annotations. 
Therefore, we denote the $i^{th}$ HPT by $\Delta_{i} = t_i^{mid} - T_i^{mid}, 0\leq i \leq N-1.$
where $N$ is the number of syllables in a sentence.
\subsubsection{Lip vowel to end}
Lip vowel to end (LVE) refers to the time difference between the lip target instant of a vowel and the sentence end.
It is observed that HPT decreases sharply as the cuer is about to finish a sentence.
This phenomenon can be explained from the perspective of cognitive function in the brain.
The articulatory process of pronunciation lags behind the hand gestures during the expression, but as the expression is nearing completion, the two streams will gradually approach synchronization. 
Therefore, we define the variable that measures the time difference between the sentence end and the lip target instant of a vowel segment  as lip vowel to end (LVE), denoted by $\delta_i$.
The mapping relationship from $\delta_i$ to $\Delta_{i}$ is written as $\hat{\Delta}_i = f_0(\delta_i)$.

\subsubsection{Lip vowel interval}
Pauses within an utterance may play a similar role for HPT at the end of the sentence.
In other words, HPT decreases sharply to a small value when there is a large time gap between two syllables.
Besides, a Mandarin syllable may have zero, one or two consonants preceding its vowel.
When the hand moves from a vowel position to the next, a hand-shape transition occurs during Mandarin CS production  if it is decorated with two consonants.
The transition can increase the time duration between two syllables, thus lengthening the HPT.
Therefore, we define the variable that measures the time difference between two adjacent lip target instants as lip vowel interval (LVI), denoted by $\alpha_i$, i.e. $\alpha_i = t_i^{mid}-t_{i-1}^{mid}$.
The last LVI of a sentence is set to the maximum value among previous LVI, i.e. $\mathop{max}\limits_{0<i<N-1} \alpha_i$.
The mapping relationship from $\alpha_i$ to $\Delta_{i}$ is written as $\hat{\Delta}_i = f_1(\alpha_i)$.

\subsubsection{Lip vowel duration}
Another factor that may influence $\Delta_i$ is the amount of time that is spent pronouncing the vowel.
Since half of the vowel segment is included in the HPT measurement, if the vowel is pronounced longer, HPT will be prolonged accordingly. 
Thereby, we refer to this measurement as lip vowel duration (LVD) and denote it by $\beta_i$.
The mapping from $\beta_i$ to $\Delta_i$ is denoted by $\hat{\Delta}_i=f_2(\beta_i)$.

\subsection{Cuer normalization}
 
 The corpus adopted in this work consists of five cuers with different ages, genders, hearing conditions, and proficiencies of CS production. 
Besides, the number of words and the expression speed vary substantially across different videos.
Hence, selecting appropriate normalization is necessary in order to address subject differences, i.e. the issue of cuer adaptation.
\subsubsection{Z-score normalization}
Since the statistical distributions of HPT and LVD are approximately normal distributions, Z-score normalization is applied.
The normalized variables are denoted by $\Delta_i'$ and $\beta_i'$, respectively.
The descriptive statistics in the unit of milliseconds for each cuer, average on all cuers, average on normal cuers, and average on deaf cuers are shown in Table \ref{tab:cuer_stat}.
The total number of vowel segments is listed in the last column of the table.

\begin{table}[th]
\footnotesize
  \caption{Descriptive statistics of the data set}
  \label{tab:cuer_stat}
  \centering
  \small
  \begin{tabular}{c|cc|cc|cc}
    \toprule
    \hline
        \multirow{2}{2.6pc}{\textbf{Cuers}} 
        & \multicolumn{2}{c|}{\textbf{HPT}}   
        & \multicolumn{2}{c|}{\textbf{LVD}} 
        & \multirow{2}{3pc}{\textbf{Syllable number}} \\ 
        \cline{2-5} 
        & \multicolumn{1}{c}{$\mu_{\Delta}$} & $\sigma_{\Delta}$ & \multicolumn{1}{c}{$\mu_{\beta}$} & $\sigma_{\beta}$   \\
        \hline
        NF1 & \multicolumn{1}{c}{242} & 177 & \multicolumn{1}{c}{338} & 79  & 10556 \\ \hline 
        NF2 & \multicolumn{1}{c}{246} & 150 & \multicolumn{1}{c}{389} & 94 & 10556 \\ \hline
        NM1 & \multicolumn{1}{c}{352} & 137 & \multicolumn{1}{c}{464} & 144   & 10556 \\ \hline
        DF1 & \multicolumn{1}{c}{154} & 110 & \multicolumn{1}{c}{365} & 97   & 4584 \\ \hline
        DM1 & \multicolumn{1}{c}{164} & 105 & \multicolumn{1}{c}{397} & 101  & 4440 \\  \hline \hline
        NORMAL & \multicolumn{1}{c}{280} & 164 & \multicolumn{1}{c}{397} & 121 & 31668 \\  \hline
        DEAF & \multicolumn{1}{c}{159} & 108 & \multicolumn{1}{c}{369} & 99  & 9024 \\  \hline
        ALL & \multicolumn{1}{c}{253} & 161 & \multicolumn{1}{c}{390} & 117  & 40692 \\  
        \hline
    \bottomrule
  \end{tabular}\\
  \footnotesize{\textit{*N/D stands for normal/deaf; F/M stands for female/male. NF1 represents ``Normal Female 1", NF2 represents ``Normal Female 2", and so forth. NORMAL refers to the corpus subset containing NF1, NF2 and NM1. DEAF refers to the corpus subset containing DF1 and DM1. ALL refers to the corpus containing five cuers. }}
\vspace{-0.5cm}
\end{table}

\subsubsection{Log-scale normalization}
Log-scale normalization is used for LVI as it shows a left-skewed distribution, denoted by $\alpha_i'=\lg \alpha_i$.
Due to the different lengths and number of characters among videos, the statistical distribution of LVE is also skewed to left.
So log-scale normalization is suggested for LVE as well, denoted by $\delta_i'=\lg \delta_i$.

\begin{figure}
\centering
\begin{subfigure}{0.15\textwidth}
\centering
\includegraphics[width=\textwidth]{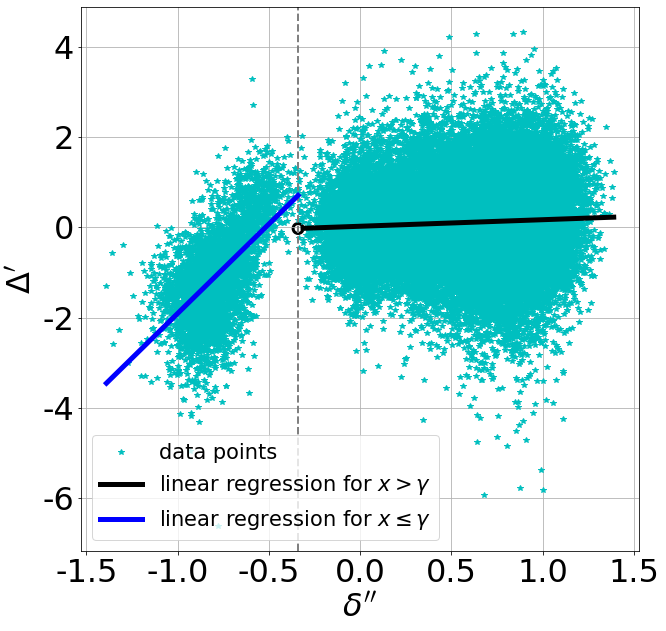}
\caption{$f_0(\delta')$}
\end{subfigure}
\begin{subfigure}{0.15\textwidth}
\centering
\includegraphics[width=\textwidth]{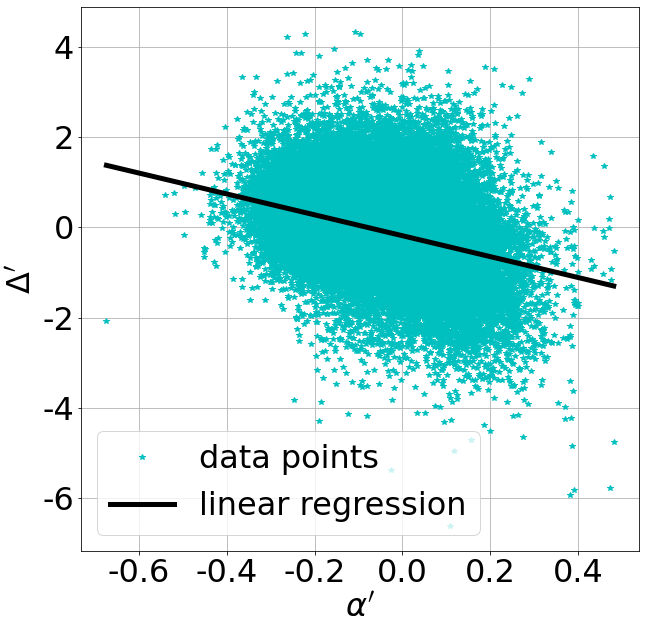}
\caption{$f_1(\alpha')$}
\end{subfigure}
\hfill
\begin{subfigure}{0.15\textwidth}
\centering
\includegraphics[width=\textwidth]{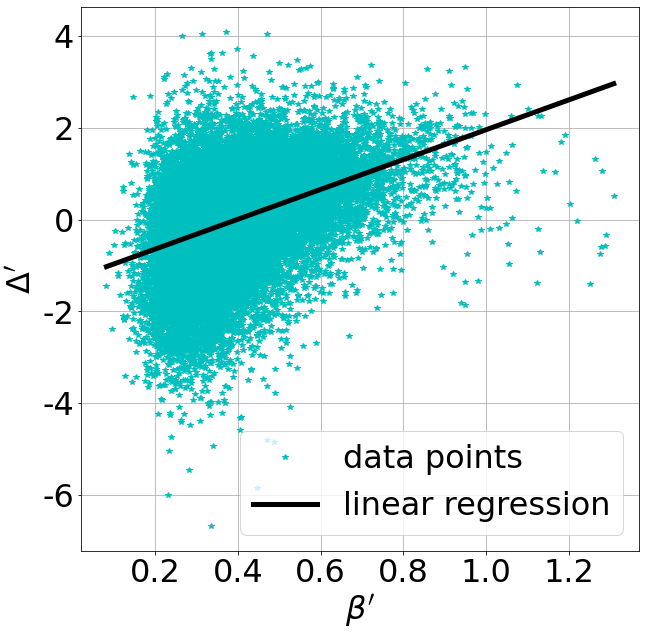}
\caption{$f_2(\beta')$}
\end{subfigure}
\caption{Linear regressions of HPT with respect to (a) LVE, (b) LVI and (c) LVD in the normalized scales.}
\label{fig:fitting}
\vspace{-0.5cm}
\end{figure}
\subsection{Temporal organization}
Fig.\ref{fig:fitting} shows the linear regressions (LRs) \cite{su2012linear} $f_0(\delta')$, $f_1(\alpha')$ and $f_2(\beta')$ based on ALL. 
Due to the presence of a distinct inflection point towards the end of the sentence, a piece-wise LR is employed for $f_0$, where the segment point is denoted by $\gamma$. 
It is observed that HPT is dominated by LVE when $\delta'$ is smaller than $\gamma$.
Otherwise, when the lip target instant is away from the sentence end, we propose a linear combination of $f_1(\alpha')$ and $f_2(\beta')$ to predict HPT. As in Eq.(\ref{eq:f}), $f(.)$ is a segmentation function taking in the normalized LVE, LVI, and LVD as input variables to calculate HPT, where $\lambda_1+\lambda_2=1$ and $\lambda_1>0, \lambda_2>0$.
Intuitively, If LVI is larger for a syllable in a sentence, it is logical to assume that $f_1(\alpha')$ carries a greater weight and the same holds true when LVD is larger.
Therefore we define $\lambda_1$ and $\lambda_2$ as in Eq. (\ref{eq:lambda}), where $\bar{\alpha}$ and $\bar{\beta}$ are the mean values of $\alpha$ and $\beta$ in a sentence, respectively.
\begin{equation}
f(\delta', \alpha', \beta') = \left\{
\begin{array}{c}
f_0(\delta'), \quad \delta' \leq \gamma \\
\lambda_{1} f_{1}(\alpha')+\lambda_{2} f_{2}(\beta'), \quad \delta'>\gamma 
\end{array}\right. 
\label{eq:f}
\end{equation}
\begin{equation}
\lambda_{1} = \frac{\alpha}{\alpha+C \beta}, \lambda_{2} = \frac{\mathrm{C} \beta}{\alpha+C \beta}, C = \frac{\bar{\alpha}}{\bar{\beta}}.
\label{eq:lambda}
\end{equation}
In the following experiments, $\gamma$ is empirically set to $-0.34$ and LR analyses are conducted for ALL, NORMAL and DEAF separately, named A-LR, N-LR and D-LR for short, respectively. 

\section{Experimental Setup}

\subsection{Data pre-processing}

To evaluate the effectiveness of the proposed HPT model, we conduct the vowel recognition experiment and calculate the average distance between the hand coordinates at the target time point and the predicted time point. 
Therefore, video clips containing vowels are segmented based on the lip annotations, hand annotations and the predicted hand target instants. The regions of interest (ROIs) of lip and hand are then extracted from each clip using the facial landmark detector included in the dlib library \cite{king2009dlib,king_dlib-models_2020} and hands solution in mediapipe \cite{mediapipehands}, respectively.
Once the key points of the lips and hands have been extracted, a 140$\times$140 square is obtained around the lips, while a 320$\times$320 square is obtained around the hands, with the exception of DM1, whose hand square is set to 400$\times$400 due to larger hand size.
To do the vowel recognition, the training and testing sets are divided randomly in a 4:1 ratio based on sentences.

\begin{table*}[t] 
\footnotesize
  \caption{The results of vowel recognition.}
  \label{tab:result1}
  \centering
  \begin{tabular}{c|c|cc|ccccc}
    \toprule
        \multirow{2}*{\textbf{LR data}} &
        \multirow{2}*{\textbf{Cuers}}
            & \multicolumn{2}{c|}{\textbf{Mono-Modal}}
            & \multicolumn{5}{c}{\textbf{Multi-Modal Fusion (Lip + Hand)}} \\ \cline{3-9}
            & {~} & Lip & Hand & Ground truth & Audio-based & Mean-based \cite{liu2020re}& LVE & LVE + LVI + LVD   \\
            \hline
        \multirow{6}{*} {ALL} &
         NF1 & \multicolumn{1}{c}{86.52\%} & 81.19\% & \multicolumn{1}{c}{97.02\%} & 89.92\% & 95.87\% & 96.35\% & 96.50\% \\
        \cline{2-9}
        {} & NF2 & \multicolumn{1}{c}{93.91\%} & 77.16\% & \multicolumn{1}{c}{98.27\%} & 96.31\% & 98.27\% & 98.27\% & 98.39\%  \\
        \cline{2-9} 
        {} & NM1 & \multicolumn{1}{c}{81.57\%} & 71.79\% & \multicolumn{1}{c}{96.40\%} & 84.69\% & 95.11\% & 95.30\% & 95.30\%  \\
        \cline{2-9}
        {} & DF1 & \multicolumn{1}{c}{86.20\%} & 80.54\% & \multicolumn{1}{c}{96.84\%} & 91.96\% & 96.09\% & 96.09\% & 96.41\%  \\
        \cline{2-9}
        {} & DM1 & \multicolumn{1}{c}{76.59\%} & 73.18\% & \multicolumn{1}{c}{91.29\%} & 85.76\% & 90.59\% & 90.71\% & 90.82\%  \\
        \cline{2-9}
        {} & Average & \multicolumn{1}{c}{86.06\%} & 76.78\% & \multicolumn{1}{c}{96.56\%} & 90.01\% & 95.76\% & 95.95\% & 96.06\% \\
        \hline 
        NORMAL & Average & \multicolumn{1}{c}{87.33\%} & 76.71\% & \multicolumn{1}{c}{97.23\%} & 90.31\% & 96.42\% & 96.51\% & 96.61\% \\
        \hline 
        DEAF & Average & \multicolumn{1}{c}{81.58\%} & 77.01\% & \multicolumn{1}{c}{94.18\%} & 88.98\% & 93.45\% & 93.56\% & 93.90\% \\
    \bottomrule
  \end{tabular}\\
  \footnotesize{\textit{*N/D stands for normal/deaf; F/M stands for female/male. NF1 represents ``Normal Female 1", NF2 represents ``Normal Female 2", and so forth.}}
\end{table*}

\subsection{Vowel recognition}
We utilize the Slowfast video classification network \cite{feichtenhofer2019slowfast} as the backbone for feature extraction and concatenate the lip and hand features for multi-modal feature fusion. The fused feature is then classified using a two-layered multilayer perceptron into 16 vowel categories.

The network is initialized randomly and trained using the SGD optimizer with an initial learning rate of 0.01, which is adjusted using the CosineAnnealing policy \cite{2016SGDR}. The mini-batch size is set to 64, and the entire network is trained for 100 epochs.


\subsection{Metrics}
Three metrics are adopted to evaluate the performance of HPT predictions. The first is mean squared error (MSE) between the HPT predictions and their ground truth in the normalized scale, denoted by $e_{\rm HPT}$. The second is mean hand coordinate distances (MHCD) between the hand coordinates at $T_i^{mid}$ and $t_i^{mid}-\hat{\Delta}_i$, denoted by $d_{\rm HPT}= \frac{1}{N}\sum_{i=0}^{N-1} \sqrt{(x_i-\hat{x}_i)^2 + (y_i-\hat{y}_i)^2}$, where $(x_i, y_i)$ and $(\hat{x}_i,\hat{y}_i)$ are the hand coordinates\footnote{The hand coordinates are obtained either by taking the fingertip or by taking the midpoint between the index finger and the middle finger. When one of the two fingers is not extended, the only extended finger (either the middle finger for hand shape 3 or the index finger for hand shapes 1 and 6) is taken.} of two time instants.
 The third is the vowel recognition accuracy, defined as the ratio of correctly recognized vowels to the total number of vowels.

\section{Results and Discussion}
The hand temporal segmentation is done through five HPT calculations: ground truth by hand annotations, named GT; baseline by audio annotations, named audio-based; mean HPT value of the dataset, named mean-based; HPT by $f_0(\delta')$, named LVE; HPT by Eq.(\ref{eq:f}), named LVE + LVI + LVD.
Here, the other combinations such as LVE + LVI and LVE + LVD are omitted since they exhibited inferior results compared to the selected ones.

\subsection{Evaluation}
\begin{figure}[t]
    \begin{subfigure}{0.28\textwidth}
    \centering
    \includegraphics[width=\textwidth]{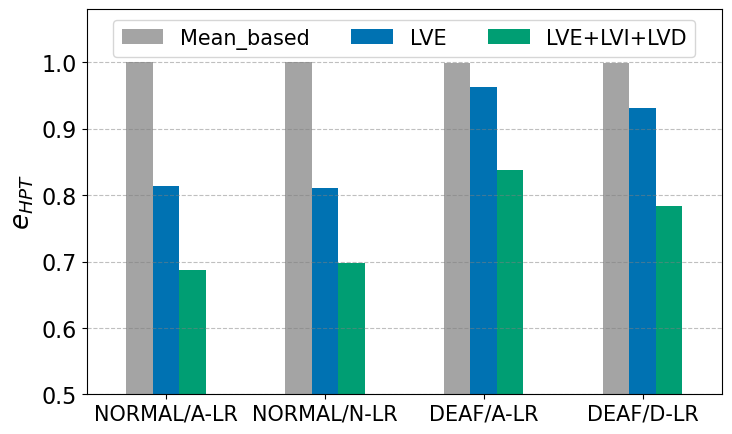}
    \caption{MSE}
    \end{subfigure}
    \centering 
    \begin{subfigure}{0.17\textwidth}
    \centering
    \includegraphics[width=\textwidth]{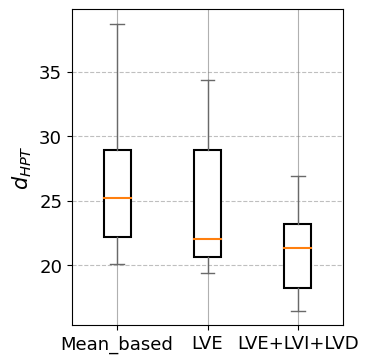}
    \caption{MHCD}
    \end{subfigure}
    \caption{Comparison of MSE and MHCD: (a) MSE across different corpus subsets calculated on different LRs. (b) Boxplot of MHCD across different cuers calculated on different LRs.} 
    \label{fig:mse} 
\vspace{-0.4cm}
\end{figure}

The vowel recognition results are shown in Table \ref{tab:result1}. 
LVE + LVI + LVD achieves the closest accuracy to GT. It even exceeds GT for NF2 by A-LR, reaching as much as 98.39\% accuracy rate, which may be due to inaccurate manual annotations.
The results of MSE and MHCD are shown in Fig. \ref{fig:mse}.
Comparing with mean-based, LVE performs better on both metrics, while LVE + LVI + LVD leads to the minimum MSE and MHCD. Besides, Fig. \ref{fig:mse}(b) shows that LVE + LVI + LVD has better stability.

\begin{figure}[t]
    \begin{subfigure}{0.23\textwidth}
    \centering
    \includegraphics[width=\textwidth]{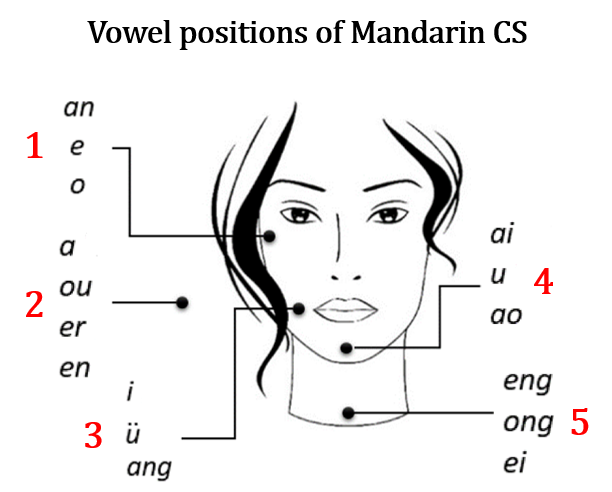}
    \caption{Vowel positions}
    \end{subfigure}
    \centering 
    \begin{subfigure}{0.23\textwidth}
    \centering
    \includegraphics[width=\textwidth]{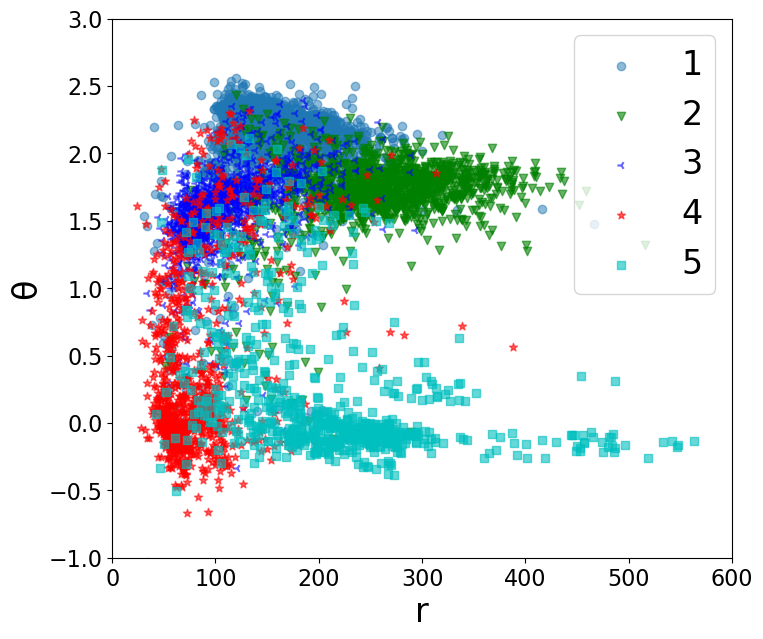}
    \caption{Audio-based}
    \end{subfigure}
    \begin{subfigure}{0.23\textwidth}
    \centering
    \includegraphics[width=\textwidth]{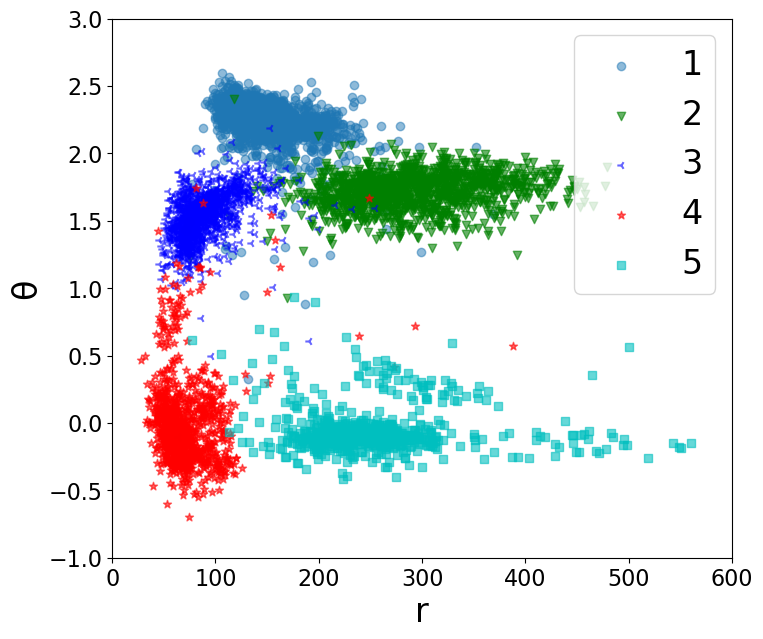}
    \caption{LVE + LVI + LVD}
    \end{subfigure}
    \begin{subfigure}{0.23\textwidth}
    \centering
    \includegraphics[width=\textwidth]{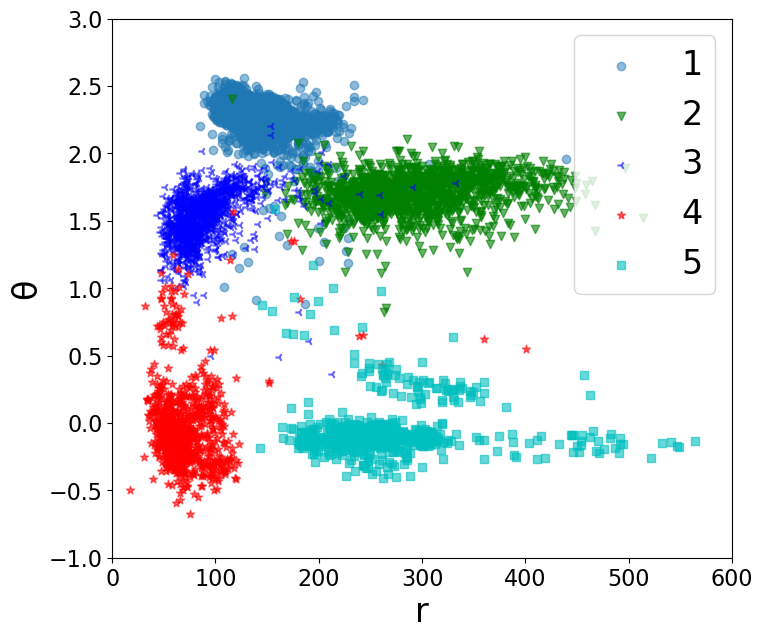}
    \caption{Ground truth}
    \end{subfigure}
    \caption{(a) Five hand positions and their corresponding vowels of the Mandarin CS system. (b)-(d) Spatial distributions of hand positions on the polar coordinates at $t_i-\hat{\Delta}_i$ of the vowels randomly selected from five cuers .} 
    \label{fig:hand} 
\vspace{-0.5cm}
\end{figure}

To visualize the correcting effect of the method, polar coordinates are used to represent hand position, where the center point of the lip is set to be the coordinates origin. The spatial distributions of hand positions at the predicted hand target instants are shown in Fig. \ref{fig:hand}. LVE + LVI + LVD has a corrective effect on the hand positions at target instants by revising the HPT values. Intuitively, a better hand temporal segment corresponds to a more separable hand position distribution. It is obvious that the separability across five positions becomes more and more remarkable from (b) to (d). 
In summary, the proposed method can improve the performance of hand temporal segmentation by comprehensive evaluation.

\subsection{Comparison between normal and hearing-impaired}
As shown in Table \ref{tab:cuer_stat}, the statistic characteristics are considerably different between NORMAL and DEAF.
A noteworthy observation is that the mean HPT of DEAF is significantly lower than that of NORMAL. 
This may be because hearing-impaired subjects have less accurate speech perception, leading to their hand movements arriving closer in time to the lip target instant.
Moreover, female cuers generally exhibit smaller mean HPT values than male cuers, indicating that women's language proficiency extends to cued speech.

In Fig. \ref{fig:mse}(a), the MSE of NORMAL by A-LR are almost the same as NORMAL by N-LR. 
But the MSE reduction in DEAF by D-LR is significant. This is reasonable as the amount of DEAF data occupies less than 25\% in the complete corpus, so that there is a larger bias compared with the NORMAL group.
This finding also gives a hint that it may be better to distinguish between different hearing conditions when developing a CS recognition system.

\section{Conclusions}
We propose a novel re-synchronization method for the lip-hand asynchronous problem during CS production.
A multi-cuer mandarin CS corpus is collected to build the statistical-based model.
Extensive experiments illustrate the efficiency of our proposed method and the differences that exist between normal and hearing-impaired cuers are discussed.
In the future, in addition to expanding the corpus, we will use the interpretable model to provide semi-supervised signals to train a pre-trained model for automatic CS segmentation and recognition.



\bibliographystyle{IEEEtran}
\bibliography{main}

\end{document}